\documentclass[10pt,aps,prd,preprint,superscriptaddress,showpacs,nofootinbib,floatfix,amsmath,amssymb]{revtex4-1}

\usepackage{amsmath}
\usepackage{amsfonts}
\usepackage{amssymb}
\usepackage{mathrsfs}

\usepackage{color}
\usepackage{hyperref}

\DeclareMathOperator{\tr}{tr}
\DeclareMathOperator{\Vol}{Vol}

\def\B{\mathcal{B}}
\def\D{\mathcal{D}}
\def\F{\mathcal{F}}
\def\M{\mathcal{M}}

\begin{document}

\title{Geometric entropy and edge modes of the electromagnetic field}

\author{William Donnelly}
\affiliation{
Department of Physics, \\
University of California, Santa Barbara \\
Santa Barbara, California 93106, USA
}
\email{williamdonnelly@gmail.com}
\author{Aron C. Wall}
\affiliation{
School of Natural Sciences, \\
Institute for Advanced Study \\ 
Princeton, New Jersey 08540, USA}
\email{aroncwall@gmail.com}

\begin{abstract}
We calculate the vacuum entanglement entropy of Maxwell theory in a class of curved spacetimes by Kaluza-Klein reduction of the theory onto a two-dimensional base manifold.
Using two-dimensional duality, we express the geometric entropy of the electromagnetic field as the entropy of a tower of scalar fields, constant electric and magnetic fluxes, and a contact term, whose leading order divergence was discovered by Kabat.
The complete contact term takes the form of one negative scalar degree of freedom confined to the entangling surface.
We show that the geometric entropy agrees with a statistical definition of entanglement entropy that includes edge modes: classical solutions determined by their boundary values on the entangling surface.
This resolves a longstanding puzzle about the statistical interpretation of the contact term in the entanglement entropy.  We discuss the implications of this negative term for black hole thermodynamics and the renormalization of Newton's constant.
\end{abstract}

\maketitle

\tableofcontents

\newpage

\section{Introduction}

The entanglement entropy of a region of space \cite{Sorkin1983,Bombelli1986,Srednicki1993} is a quantity with broad applications including to black hole physics \cite{Solodukhin2011}, condensed matter theory \cite{Amico2007} and the AdS/CFT correspondence \cite{Ryu2006a,Nishioka2009}.
In all of these applications one encounters field theories with gauge symmetry, and for gauge theories multiple new subtleties arise that are not present in the case of scalar and spinor fields.
For minimally coupled scalars and for spinors, the entanglement entropy can be computed by Euclidean methods. For nonminimally coupled scalars, gauge fields, and gravitons the Euclidean formula contains a contact term that does not have a known interpretation as entanglement entropy \cite{Kabat1995,Fursaev1996,Solodukhin2015}.
Understanding these contact terms has been identified as one of the major open problems in the entanglement entropy of black holes \cite{Solodukhin2011}.
The goal of the present paper is to resolve these issues in the context of $U(1)$ Maxwell theory (i.e. compact QED with no charges).

The geometric entropy of a static spacetime with a bifurcate Killing horizon (such as the Schwarzschild black hole, Rindler, or de Sitter) can be calculated by means of a conical variation of the Euclidean path integral. 
In terms of the covariant partition function $Z$, the \emph{geometric entropy} is given by \cite{Callan1994}
\begin{equation}\label{conical}
S = (1 - \beta \partial_\beta) \ln Z |_{\beta = 2 \pi},
\end{equation}
where the variation of the angular period $\beta$ not only changes the temperature, but also inserts a conical singularity at the bifurcation surface of the Killing horizon.
Formally, this is similar to a protocol used to calculate the entanglement entropy, 
\begin{equation}
S(\rho) = -\tr (\rho \ln \rho),
\end{equation} 
of the reduced density matrix $\rho$ of a region bounded by an entangling surface, where in this case the entangling surface is the bifurcation surface.
While the term \emph{geometric entropy} has sometimes been used interchangeably with the term \emph{entanglement entropy}, here we wish to draw a distinction between the two quantities, as in general they can be different. 
If the fields couple nontrivially to curvature, the geometric entropy contains a contact term due to interaction of the fields with the conical singularity.
These contact terms appear for nonminimally coupled scalar fields, gauge fields, and gravitons \cite{Solodukhin2011,Solodukhin2015}.
Such contact terms need not have an interpretation in terms of a von Neumann entropy.

We will show that in the case of Maxwell theory, the contact term in fact does have a statistical interpretation: it is the entanglement entropy of edge modes, which are degrees of freedom localized on the entangling surface.

To see how contact terms arise, a useful illustrative example is that of a nonminimally coupled scalar field \cite{Larsen1995,Solodukhin1995,Kabat1995b}.
Consider a scalar field $\phi$ with the Euclidean action
\begin{equation} \label{scalar-action}
I[\phi] = \frac12 \int \sqrt{g} (\nabla_a \phi \nabla^a \phi + \xi R \phi^2).
\end{equation}
The action contains a direct coupling to curvature, which leads to a contact interaction with the conical singularity.
The contribution of this interaction to the geometric entropy takes the form of a quantum expectation value of Wald's entropy formula \cite{Wald1993,Iyer1994,Iyer1995}, which for this action takes the form:
\begin{equation} \label{scalar-contact}
S_\text{contact} = - 2 \pi \xi \int \langle \phi^2 \rangle
\end{equation}
with the integral taken over the entangling surface.
This expectation value is divergent and can have either sign depending on the parameter $\xi$ of the nonminimal coupling.  This term cannot be part of the entanglement entropy, as it would lead to the conclusion that the entanglement entropy in flat spacetime depends on the value of the nonminimal coupling parameter $\xi$. 
But the entanglement entropy should only be a function of the state $\rho$, and the vacuum wavefunction in flat spacetime is independent of $\xi$.
In the case of the nonminimally coupled scalar, the contact term can be understood as an additional contribution to the generalized entropy that must be included \emph{even classically} in order to obtain a quantity obeying the generalized second law \cite{Ford2000}.
This gives a consistent picture of the contact term for nonminimally coupled scalars, albeit one without a statistical interpretation at low energies.  However, such terms may still arise from a high energy theory in which all entropy is statistical \cite{Kabat1995b}.

The geometric entropy of Maxwell theory was first calculated by Kabat \cite{Kabat1995}.
He found that the partition function of Maxwell theory also has a contact term which can be traced to the nonminimal coupling in the spin-1 Laplacian $\Delta_1 = -g_{ab} \nabla^2 + R_{ab}$.  This contact term contributes negatively to the entropy, leading to an overall negative sign of the leading order divergence for $D < 8$.
The meaning of the contact term of Maxwell theory has remained obscure, and there has been much disagreement as to whether it should be regarded as physical \cite{Barvinsky1995,Iellici1996,Cognola1997,Kabat2012,Donnelly2012,Solodukhin2012,Eling2013,Huang2014a}.

While it may be tempting to also interpret the Maxwell contact term as a Wald entropy, this interpretation is untenable for several reasons \cite{Donnelly2012}. 
First, the coefficient of the nonminimal coupling in Maxwell theory is fixed; thus one cannot rule out an entanglement interpretation by comparing different nonminimal couplings, as was done in the case of a scalar field.
Moreover, if one repeats the argument leading to \eqref{scalar-contact} one arrives at the integral of the gauge-dependent expression $- \pi \langle A_\perp^2 \rangle$, where $A_\perp$ is the projection of the gauge potential onto to the normal plane of the entangling surface\footnote{In the family of 't Hooft gauges, the regulators can be adjusted to make the result independent of the gauge parameter $\xi$ \cite{Solodukhin2012}. 
This suggests that the contact term may contain some universal gauge-invariant information, and indeed we will show that this is the case.
However the expression in terms of fluctuations of $A_\perp$ does nothing to establish its meaning as a gauge-invariant statistical entropy.
}.
This would-be contact term in fact represents an ambiguity in the definition of Wald's entropy formula \cite{Jacobson1993,Iyer1994} when generalized to the situation of fluctuating quantum fields.
This ambiguity was recently resolved for the case of classical higher derivative gravity \cite{Dong2013,Camps2013} and when applied to the case of classical Maxwell theory, this refinement of Wald's formula gives zero \cite{DongPrivate,Huang2014a}.
Finally, unlike the nonminimally coupled scalar, there is no need to add an additional term to the generalized entropy: since Maxwell fields satisfy the null energy condition, the classical second law is already satisfied.
We take this as evidence that the divergent contact term in Maxwell theory cannot be interpreted as a quantum version of Wald entropy.
Hence some other explanation for the contact term divergence is needed.

As an alternative to the Euclidean path integral, one can calculate entanglement entropy using a physical regulator, such as a lattice \cite{Buividovich2008b,Donnelly2011,Casini2013,Casini2014,Donnelly2014a}.
In Hamiltonian lattice gauge theory (without matter), the configuration space degrees of freedom are integrals of the gauge field along links of a spatial lattice. 
The space of physical states is not the full tensor product of link Hilbert spaces, it is a quotient of this space by gauge transformations.
This physical Hilbert space thus does not admit a canonical factorization according to regions of space.
One approach \cite{Buividovich2008b,Donnelly2011} is to embed the physical Hilbert space into a tensor product of local Hilbert spaces.
The local Hilbert spaces include \emph{edge mode} degrees of freedom living on the boundary, which arise due to the Gauss constraint, that give a positive contribution to the entropy.
\footnote{This definition is closely related to the ``electric'' definition of entanglement entropy in Ref.~\cite{Casini2013} although there are some differences in topologically nontrivial regions.}
In the case of the toric code, this definition reproduces the well-known value for the universal subleading term in the entanglement entropy (the topological entanglement entropy \cite{Kitaev2005,Levin2005}) which persists in the continuum limit.
In this case the entire entropy, including its universal piece, comes from the sum over edge modes.
So the edge modes are essential for obtaining the universal terms in the continuum entanglement entropy, and we will see that the same is true for Maxwell theory.

In Ref.~\cite{Donnelly2012}, the contact term was studied with a focus on the case of two-dimensional spacetimes, particularly those such as two-dimensional de Sitter which are compact after Wick rotation.
There it was found that once the topological sector of the theory is treated correctly, the geometric entropy is equal to the entanglement entropy.
However two dimensions is a rather special case, since two-dimensional Maxwell theory has only global degrees of freedom.

The goal of the present paper, which expands on the arguments of Ref.~\cite{Donnelly2014b} (cf. \cite{Huang2014b}), is to extend the analysis of Ref.~\cite{Donnelly2012} to spacetime dimension $D > 2$, using a continuum analogue of the lattice entropy defined in Refs.~\cite{Buividovich2008b,Donnelly2011}.
We will use a result for the partition function of Maxwell theory that properly takes into account the effects of the compact gauge group \cite{Donnelly2013}, which is reviewed in section \ref{sec:max}.

We consider product manifolds of the form $\B \times \F$, where $\B$ is a two-dimensional manifold with a bifurcate Killing horizon and a compact Euclidean section (the \emph{base}), and $\F$ is any compact manifold (the \emph{fiber}).
For example, we can consider a geometry in which one spatial dimension is exponentially expanding to the past or future, while the other dimensions stay a fixed size.
This is represented by a geometry $dS_2 \times \F$, which Wick rotates to $S^2 \times \F$.

We then treat the contact term by Kaluza-Klein reducing onto $\B$ in section \ref{sec:kk}.
The U(1) Maxwell theory on the manifold $\B \times \F$ reduces to multiple U(1) Maxwell theories on $\B$ (representing electric and magnetic fluxes), together with a number of periodic massless scalar fields and towers of massive scalar and vector fields.
The advantage of this reduction is that we can dualize all vector degrees of freedom on $\B$ to scalar degrees of freedom.
This leaves us with towers of fields for which the geometric entropy agrees with the entanglement entropy; what remains is the contact term.
This contact term takes the form of a negative scalar field confined to the entangling surface.
Its leading divergence agrees with the result of Ref.~\cite{Kabat1995}, but we also establish the existence of subleading and finite terms, some of which are universal, i.e. independent of the regulator scheme.


In order to give a statistical interpretation to the contact term, we consider regulating the conical singularity by introducing a ``brick wall'' \cite{tHooft1985} at a short distance $\epsilon$ from the entangling surface.
When standard boundary conditions are fixed at the brick wall, the geometric entropy formula has a statistical interpretation as the entropy of a thermal ensemble with fixed boundary conditions.
However, the brick wall does not capture the correct physics of the entangling surface, which does not obey any boundary conditions.
In section \ref{sec:geo}, we discuss how the partition function changes under the introduction of a brick wall.
For magnetic conductor boundary conditions, the partition function is the same as if there was no brick wall, except for a small correction coming from exchanging Dirichlet with Neumann boundary conditions on some of the scalar fields (calculated in section \ref{sec:dn}), and an edge mode contribution.


In section \ref{sec:edge} we explain more carefully the origin of the edge modes, and calculate their partition function.  This allows us to confirm that the geometric entropy agrees with the statistical entropy.
In particular, the contact term captures the entanglement entropy of the edge modes.


In section \ref{sec:log} we consider the case of four-dimensional spacetime.
Four dimensions is special because Maxwell theory is conformal, and so the logarithmic divergence of the entanglement entropy is universal and should be related to the trace anomaly \cite{Solodukhin2008,Casini2011}.
But the trace anomaly result was found to be in conflict with the entanglement entropy calculated by thermodynamic methods \cite{Dowker2010,Eling2013,Huang2014b}.
We resolve this puzzle by showing that when the edge modes are included, the entanglement entropy agrees with the trace anomaly.  We comment on the implications for the holographic entanglement entropy at strong coupling, which we argue must already contain an edge mode contribution from the strongly coupled Yang-Mills theory.

In the Discussion, we explain why the leading order contribution found in Ref.~\cite{Kabat1995} is negative, and explain how the sign of the leading order term depends on the choice of cutoff.  We also discuss possible extensions of our work to nonabelian gauge fields and gravitons, or to entangling surfaces without Killing symmetry.  Finally we discuss the implications for black hole physics and the renormalization of Newton's constant.

\section{Maxwell theory} \label{sec:max}

To prepare for Kaluza-Klein reduction, we first consider the partition function of Maxwell theory with gauge group U(1) on a compact Euclidean manifold $\M$.

Locally the electromagnetic field can be represented as a $1$-form $A$ up to local gauge transformations $A_a \to A_a + \nabla_a \alpha$ where $\alpha$ is a scalar.
The Euclidean action $I$ is expressed in terms of the electromagnetic field tensor $F_{ab} = \nabla_a A_b - \nabla_b A_a$ as
\begin{equation}
I[F] = \int d^4x \sqrt{g}\, \frac{1}{4} F^{ab} F_{ab}.
\end{equation}
The partition function is then given formally by the Euclidean path integral 
\begin{equation}
Z = \frac{1}{\Vol(G)} \int \D A \; e^{-I[dA]}.
\end{equation}
There are global issues that arise from the $U(1)$ nature of the gauge field:

First, we must identify any two $1$-forms $A$ and $A'$ such that around every closed curve $\gamma$
\begin{equation} \label{largegauge}
\oint_\gamma A - \oint_\gamma A' \in \frac{2 \pi}{q} \mathbb{Z}.
\end{equation}
This requirement ensures that a particle whose charge is a multiple of $q$ cannot distinguish $A$ from $A'$ when transported around a noncontractible curve.
Equivalently, one can allow for the parameter $\alpha$ of the gauge transformation to be identified under $\alpha \to \alpha + 2 \pi/q$ when going around a noncontractible curve.  These are the large gauge transformations.

Second, we must include field strength tensors $F$ that can be expressed as $F = dA$ locally, but globally require gluing multiple $A$ fields together with the requirement that $\oint A$ agrees up to a multiple of $2 \pi / q$ where the patches overlap.
This leads to the Dirac quantization condition, $\oint_\Omega F \in 2 \pi / q$ for every closed 2-surface $\Omega$.
The set of such field configurations is discrete, so they are summed over in the path integral.

Though for many purposes one does not need to distinguish between the gauge groups $U(1)$ and $\mathbb{R}$, here the distinction is fundamentally important.
The reason is that while the nonzero modes of a gauge field act like harmonic oscillators, the zero modes of the $\mathbb{R}$ gauge theory act like free particles, and do not have a normalizable ground state.
Since we are calculating the ground state entanglement entropy, a noncompact gauge group would lead to an infrared divergence in the entanglement entropy.
Attempting to cure this divergence by the introduction of a mass breaks gauge invariance and leads to further problems.
This problem is naturally cured in the $U(1)$ gauge theory, since the zero modes are quantum mechanical free particles on a circle, which again have normalizable ground states.

In Ref.~\cite{Donnelly2013}, the partition function of Maxwell theory was calculated in the Euclidean path integral by covariant gauge-fixing.
The result can be expressed as a product of terms:

First, there are functional determinants that arise from the Gaussian path integral over the nonzero modes of the vector potential and the Faddeev-Popov ghosts.
In the covariant formalism these consist of modes of the transverse vector Laplacian $\Delta_1^T$ and of the scalar Laplacian $\Delta_0$.
The longitudinal modes depend on the `t Hooft parameter $\xi$, and on a mass scale $\mu$ appearing in the path integral measure, which we have allowed to take different values for the vector field ($\mu_A$) and ghosts ($\mu_\alpha$).
The nonzero modes contribute to the partition function a factor 
\begin{equation} \label{nonzero-modes}
Z_\text{nonzero modes} =
\det{}' \left( \frac{\Delta_1^T}{2 \pi \mu_A^2} \right)^{ -1/2}
\det{}' \left( \frac{\Delta_0}{2 \pi \mu_A^2 \xi} \right)^{-1/2}
\det{}' \left( \frac{\Delta_0}{\mu_\alpha^2} \right).
\end{equation}
The prime denotes that $\det{}'$ is the product only over nonzero eigenvalues.

Second, there is a factor coming from the flat connections. 
Since the action vanishes for these configurations, they contribute a factor of the volume of their moduli space. 
This volume is made finite by the quotient by large gauge transformations.
Let $w_I = w_{Ia} dx^a$, $I = 1, \ldots, b_1$ be a topological basis of 1-forms in $H^1(\M,\mathbb{Z})$, whose dimension is the Betti number $b_1$.
Any harmonic 1-form whose integrals around all closed curves are integers can be written uniquely as an integer linear combination of the $w_I$.
In this basis, the space of flat connections modulo large gauge transformations is the torus obtained by identifying opposite edges of the cube $[0,1]^{b_1}$.
The standard norm on vector fields pulled back to this space defines the metric on moduli space as
\begin{equation}
\Gamma_{IJ} = \int_\M \sqrt{g} \, g^{ab} \, w_{Ia} w_{Jb} .
\end{equation}
The contribution of flat connections to the path integral is simply the volume of moduli space in the functional measure, and is given by
\begin{equation} \label{flat-connections}
Z_\text{flat connections} = \det \left( \left[\frac{2 \pi \mu_A}{q} \right]^2 \Gamma \right)^{1/2}.
\end{equation}

Third, there is a factor associated with the constant gauge transformations. 
Since these gauge transformations do not modify $A$ they must be treated specially; we still must divide by the volume of the gauge group, but the zero mode cannot be gauge fixed as is conventionally done for the higher modes.
The result is a global factor that depends on the volume $V$ of the spacetime manifold:
\begin{equation} \label{constant-gauge-transformations}
Z_\text{constant gauge transformation} = \frac{q}{\mu_\alpha^2} \sqrt{\frac{\xi}{2 \pi V}}.
\end{equation}
We note that this term is often absent from discussions of the path integral of gauge theories, but its presence is essential for agreement with the canonical formalism as shown in \cite{Donnelly2013}.

Finally, there is a factor associated to the nontrivial bundles.
These are classified by the (discrete) second homology group $H^2(M, \mathbb{Z})$, which consists of harmonic 2-forms whose integrals over all closed 2-surfaces are integers.
Their contribution to the partition function is
\begin{equation} \label{nontrivial-bundles}
Z_\text{nontrivial bundles} = \sum_{F \in \frac{2 \pi}{q} H^2(M, \mathbb{Z})} e^{-I[F]}.
\end{equation}

Putting all of these factors \eqref{nonzero-modes}
\eqref{flat-connections}\eqref {constant-gauge-transformations}
\eqref{nontrivial-bundles} together, the result is
\begin{equation} \label{Zlong}
Z = \frac{q}{\mu_\alpha^2} \sqrt{\frac{\xi}{2\pi V}}
\det \left( \left[\frac{2 \pi \mu_A}{q} \right]^2 \Gamma \right)^{1/2}
\det \left( \frac{\Delta_1^T}{2 \pi \mu_A^2} \right)^{-1/2}
\det \left( \frac{\Delta_0}{2 \pi \mu_A^2 \xi} \right)^{-1/2}
\det \left( \frac{\Delta_0}{\mu_\alpha^2} \right)
\sum_{F \in \frac{2 \pi}{q} H^2(M, \mathbb{Z})} e^{-S[F]}
\end{equation}

We can simplify this formula by rescaling the functional determinants using zeta function regularization:
\begin{equation}
\det{}'(\Delta) := e^{- \zeta'(\Delta,0)},
\end{equation}
where $\zeta(\Delta,s)$ is the zeta function of the operator $\Delta$, $\zeta(\Delta,s) := \tr'(\Delta^{-s})$, and $\tr'$ denotes omitting the zero modes of $\Delta$.
This formula defines the zeta function for $\text{Re}(s) > D/2$; it is then defined for other values of $s$ by analytic continuation.
We can see directly from the definition that the functional determinant scales as
\begin{equation} \label{zeta}
\det{}' (a \Delta) = a^{\zeta(\Delta,0)} \det{}'(\Delta)
\end{equation}
so that $\zeta(\Delta,0)$ can be thought of as a regularized number of nonzero modes.
We can then apply the following result, that for any elliptic differential operator,
\begin{equation}
\zeta(\Delta,0) = - \dim \ker \Delta + A
\end{equation}
where $A$ is an anomaly that appears only when $D$ is even and takes the form of the integral of a local geometric quantity.
The anomaly can be cancelled by a local counterterm; we can therefore ignore a finite shift in this term.
The factors of $\mu_A$, $\mu_\alpha$ and $\xi$ from scaling the determinants cancel with the other terms in Eq.~\eqref{Zlong}.

Thus upon rescaling the functional determinants, we see that the partition function does not depend on the measure factors $\mu$ or the gauge parameter $\xi$ (as should be the case on physical grounds):
\begin{equation} \label{Z}
Z = 
\sqrt{\frac{q^2}{2\pi \Vol(\M)}}
\det \left(\frac{2 \pi \Gamma}{q^2}\right)^{1/2}
\frac{\det{}' ( \Delta_0)}
{\det{}' (\Delta_1)^{1/2}}
\sum_{F \in \frac{2 \pi}{q} H^2(M, \mathbb{Z})} e^{-I[F]}.
\end{equation}
By scaling out the factor of $\xi$ from the longitudinal determinant we have effectively chosen Feynman gauge ($\xi = 1$), and this has allowed us to combine the longitudinal and transverse components of the vector field into a single determinant.

\section{Kaluza-Klein reduction} \label{sec:kk}

We now consider Kaluza-Klein reduction of Maxwell theory on a manifold of the product form $\M = \B \times \F$.
Here $\B$ is a two-dimensional base manifold and $\F$ is a $D-2$ dimensional compact fiber: $\B$ will contain the directions normal to the entangling surface, and $\F$ the directions along the entangling surface.
We will carry out the reduction at the level of the partition function in order to keep off-shell effects to which the entanglement entropy is sensitive.
The purpose is to divide the Maxwell partition function into a factor for which the geometric entropy formula \eqref{conical} agrees with the entanglement entropy of the on-shell degrees of freedom, and another portion that we identify as the contact term.

We will show that the Maxwell partition function can be written as a product of partition functions 
\begin{equation}
Z = Z_\text{scalars} Z_\text{E} Z_\text{B} Z_\chi
\end{equation}
where $Z_\text{scalars}$ is a tower of scalar fields on $\B$; $Z_\text{E}$ and $Z_\text{B}$ are two-dimensional Maxwell fields on the base corresponding to electric and magnetic fluxes respectively.
For these first three contributions, the geometric entropy is equal to the entanglement entropy.
The remaining factor $Z_\chi$ is the contact term, whose interpretation will be the subject of section \ref{sec:edge}.

\subsection{Unpacking the partition function} \label{subsec:unpack}
  
  
We first consider the functional determinant piece of the partition function \eqref{nonzero-modes}. 
On a product manifold $\B \times \F$ we can split the vector determinant into a contribution from vectors polarized along the base $\B$, and a contribution from vectors polarized along the fiber $\F$.
This is expressed in the identity
\begin{equation} \label{functional-determinant-identity}
\det{}' (\Delta_1) =
\det {}' (\Delta^\B_1 \oplus \Delta^\F_0) \det {}' (\Delta^\B_0 \oplus \Delta^\F_1).
\end{equation}
Viewed from the base manifold, the vector field breaks into a Kaluza-Klein tower of vector fields whose masses are given by the spectrum of the scalar Laplacian on the fiber, $m^2 \in \text{spec}(\Delta_0^\F)$, and a tower of scalar fields whose masses are given by the spectrum of the vector Laplacian on the fiber, $m^2 \in \text{spec}(\Delta_1^\F)$.
This functional determinant, together with the scalar functional determinant of the ghosts, encodes all the $(D-2)$ local bosonic degrees of freedom of the $D$-dimensional Maxwell field.
We now turn to the remaining parts of the partition function that describe the topological sector.

We can also decompose the moduli space of flat connections into fiber and base polarizations.
Letting $\Gamma^{(\M)}$ denote the metric on the space of flat connections on $\M$, we see that the metric on the product splits as
$\Gamma^{(\M)} = \Vol(\F) \Gamma^{(\B)} \oplus \Vol(\B) \Gamma^{(\F)}$, so that:
\begin{equation} \label{moduli}
\det \left(\frac{2 \pi \Gamma^{(\M)} }{q^2}\right)^{1/2} = 
\det \left(\frac{2 \pi \Gamma^{(\B)}} {q_\B^2} \right)^{1/2}
 \det \left(\frac{2 \pi \Vol(\B) \Gamma^{(\F)}}{q^2}\right)^{1/2}.
\end{equation}
Here we have defined $q_\B = q / \sqrt{\Vol(\F)}$, the fundamental charge of the two-dimensional Maxwell theory on $\B$. 

We can now express the prefactor from the gauge zero modes \eqref{Z} in terms of $q_\B$ as
\begin{equation} \label{prefactor}
\sqrt \frac{q^2}{2 \pi \Vol(\M)}
= \sqrt \frac{q_\B^2}{2 \pi \Vol(\B)},
\end{equation}
which we recognize as the gauge zero mode term for two-dimensional Maxwell theory on $\B$, with fundamental charge $q_\B$.

Since the bundles correspond to harmonic two-forms, they can be divided into three types depending on which two directions the two-form point: along the base, along the fiber, or both.
The harmonic two-forms on the two-dimensional base $\B$ can be expressed as $F_{ab} = f \epsilon_{ab}$, where $f$ is constant and $\epsilon_{ab}$ is the volume form on $\B$. 
They are quantized so that $\int_\B F  = \Vol(\B) f \in \tfrac{2\pi}{q} \mathbb{Z}$.
Their contribution to the partition function is:
\begin{equation} \label{basebundles}
\sum_{F \in \frac{2 \pi}{q} H^2(\B, \mathbb{Z})} e^{-\tfrac{1}{4}\int_\M F^2} = 
\sum_{F' \in \frac{2 \pi}{q_\B} H^2(\B, \mathbb{Z})} e^{-\tfrac14 \int_\B (F')^2}
\end{equation}
where we have defined the rescaled field tensor $F' = \sqrt{\Vol(\F)} F$.
In this form, we see that it is equal to the sum over the nontrivial bundles of Maxwell theory on $\B$ with fundamental charge $q_\B$.

The bundles pointing along the fiber can be expressed similarly as
\begin{equation} \label{fiberbundles}
\sum_{F \in \frac{2 \pi}{q} H^2(\F, \mathbb{Z})} e^{-\tfrac14 \Vol(\B) \int_\F F^2}
\end{equation}
These correspond to magnetic fields in the fiber directions that are constant along the base.

The mixed bundles that point in both base and fiber directions can be expressed in terms of a basis $w^\B_I$ of $H^1(\B, \mathbb{Z})$ and a basis $w^\F_J$ of $H^1(\F, \mathbb{Z})$.
A general mixed element of $H^2(\M, \mathbb{Z})$ then takes the form 
\begin{equation}
F = \left( \frac{2 \pi}{q} \right) m_{IJ} w^\B_I \wedge w^\F_J
\end{equation}
where $m_{IJ}$ is a matrix of integers.
The contribution of these mixed bundles to the partition function is
\begin{equation} \label{mixedbundles}
\sum_{m_{IJ}} \exp \left[- \frac{1}{2} \left( \frac{2 \pi}{q} \right)^2 \sum_{IJKL} m_{IJ} m_{KL} \Gamma^{(B)}_{IK} \Gamma^{(F)}_{JL} \right].
\end{equation}

Using these identities, we can express the original partition function of Maxwell theory \eqref{Z} as a product of field theories defined on $\B$.
However Eq.~\eqref{functional-determinant-identity} contains a tower of vector fields, each of which has a contact term.
In order to isolate this contact term, we will first trade these vector degrees of freedom for scalars.

\subsection{Proca-scalar duality} \label{subsec:proca}

The Kaluza-Klein description of Maxwell theory includes a tower of vector fields on the base manifold $\B$.
Since the vector Laplacian contains an effective nonminimal coupling to background curvature, the vector fields will include a contact term in addition to the contribution from their on-shell degrees of freedom.
In order to disentangle these two contributions we make use of massive $p$-form duality to relate the massive vector fields to dual scalar fields $\phi_\text{dual}$.

We perform a Hodge decomposition of the operator $\Delta_1^\B$, 
expressing the vector field $A_a$ as an orthogonal sum of exact, co-exact and harmonic vector fields.
In two-dimensions this takes the form $A_a = \nabla_a \phi + \epsilon_{ab} \nabla^b \psi + B_a$, where $\phi$ and $\psi$ are scalars and $B_a$ is a harmonic vector field.
In terms of spectra, this says that the spectrum of the vector Laplacian $\Delta_1^\B$ is two copies of the spectrum of the scalar Laplacian $\Delta_0^\B$, up to zero modes.
This leads to the functional determinant identity for $m > 0$:
\begin{eqnarray} \label{hodge}
\det (\Delta_1^\B + m^2) = \det (\Delta_0^\B + m^2)^2 \; m^{-2 \chi(\B)}.
\end{eqnarray}
The Euler characteristic $\chi(\B)$ comes from the difference between the number of zero modes of a vector and two scalars: a vector has $b_1$ zero modes, while a scalar has $b_0$ zero modes, and $2 b_0 - b_1 = \chi$.
In the massless sector, we do not include the zero modes in the functional determinant, so there is no Euler number correction and we have simply
\begin{equation}
\det{}' (\Delta_1^\B) = \det(\Delta_0^\B)^2.
\end{equation}

When we take the product of Eq.~\eqref{hodge} over the spectrum of Kaluza-Klein masses, we obtain the identity 
\begin{equation}
\det{}' (\Delta_1^\B \oplus \Delta_0^\F) = \det{}'(\Delta_0^\B \oplus \Delta_0^\F)^2 \; \det{}' (\Delta_0^\F)^{- \chi(\B)}.
\end{equation}
The first factor describes two scalar fields on $\B \times \F$; when we apply this identity to the Maxwell partition function it will cancel with the two Faddeev-Popov ghosts.
The remaining term takes the form of $-\chi(\B)$ scalar fields on $\F$.
We will see that this gives the contribution of the nonzero modes to the contact term.

This relation between functional determinants can be understood in more physical terms via the 2D duality between the massive vector (Proca) field and a massive scalar field.  Recall that the Proca action for a massive vector field is
\begin{equation}
I = \int \frac{1}{4} F_{ab} F^{ab} + \frac{1}{2} m^2 A_a A^a,
\end{equation}
where the mass term breaks the gauge symmetry $\delta A_a = \nabla_a \alpha$ of the massless vector field.  However, it is possible to restore the gauge symmetry by adding an additional scalar field $g$ which transforms as $\delta g = m \alpha$, so that the combination $\nabla_a g - m A_a$ is gauge invariant.  One can then write the action in the equivalent Stueckelberg form:
\begin{equation}
I = \int \frac{1}{4} F_{ab}^{\phantom{ab}2} + \frac{1}{2}(\nabla_a g - m A_a)^2,
\end{equation}
where the equivalence to the Proca form can be shown to hold (even off-shell) by gauge-fixing so that $g = 0$.

When we KK reduce the Maxwell field, the tower of massive vector fields naturally appear in this Stueckelberg form.  $A_a$ is proportional to the vector field polarized on the base, while the Stueckelberg mode $g$ is proportional to the corresponding longitudinal mode on the fiber.  These two modes are related by a gauge symmetry, which comes from reducing the higher dimensional gauge symmetry.  In Feynman gauge, $A_a$, $g$, and the ghosts all propagate independently, so each massive vector field has +1 degree of freedom, just like a scalar field.

On-shell, the Proca field is dual to a scalar field $\phi$ via the duality 
\begin{equation}\label{dualize}
F_{ab} \epsilon^{ab}/2 = m\phi.
\end{equation}
Although this duality does not make sense as a substitution into the action, it preserves the Hamiltonian and the equations of motion.

This on-shell duality explains why the partition function of these modes is equivalent to a massive scalar, up to the contact term
\begin{equation} \label{contact}
\det{}' (\Delta_0^F)^{\chi_B/2}.
\end{equation}
This term is an off-shell effect, so the duality does not know about it.  Each Proca field contributes a factor of $m^\chi$, coming from the difference in the number of zero modes between the Proca and scalar field.

\subsection{The contact term} 

Using the decomposition of the partition function \ref{subsec:unpack}, and the Proca-scalar duality \ref{subsec:proca} we will now rewrite the Maxwell partition function in a way that isolates the contact term.

After applying the Proca-scalar duality all the local degrees of freedom of Maxwell theory are expressed in terms of vector fields polarized along the fiber directions:
\begin{equation} \label{scalars}
\det{}'( \Delta_0^\B \oplus \Delta_1^\F )^{-1/2}.
\end{equation}
The factor \eqref{scalars} includes both massive and massless scalar fields on the base.  Some of the massive scalar fields, $\phi_\text{dual}$, came from dualizing Proca fields---these correspond to the modes of $\Delta_1^F$ which come from differentiating $\Delta_0^F$ modes.  The rest of the scalar modes (which we shall call $\phi_\text{KK}$) come from direct KK reduction.  For each positive eigenvalue of $\Delta_1^F$, Eq.~\eqref{scalars} gives the partition function of a massive scalar field on $\B$.

Among the $\phi_\text{KK}$ fields, there is one massless scalar field on the base for every vector zero mode on the fiber; for these Eq.~\eqref{scalars} gives only the contribution from the nonzero modes.
These massless modes are in fact periodic scalars; their full partition function consists of the zero modes of the scalar determinant \eqref{scalars}, the right factor of \eqref{moduli}, and the bundle sum \eqref{mixedbundles}.
Combining these factors we find the partition function of a linear $\sigma$-model:
\begin{equation} \label{sigma-model}
\det{}' (\Delta_0^\B)^{b_1(\F)} \det \left(\frac{2 \pi \Vol(\B) \Gamma^{(\F)}}{q^2}\right)^{1/2}
\sum_{\{m_{IJ} \} \in \mathbb{Z}} \exp \left[- \frac{1}{2} \left( \frac{2 \pi}{q} \right)^2 \sum_{I,J,K,L} m_{IJ} m_{KL} \Gamma^{(B)}_{IK} \Gamma^{(F)}_{JL} \right].
\end{equation}
The target space is, up to a prefactor, the metric on the space of flat connections on $\F$,
\begin{equation}
\left( \frac{2 \pi}{q} \right)^2 \Gamma^\F.
\end{equation} 
The functional determinant in \eqref{sigma-model} comes from the nonzero modes, the second factor is the integral over the (compact) zero mode, and the sum is over winding sectors.

Next we consider the massless vector modes on $\B$.
Combining the moduli space of the base \eqref{moduli}, the volume of the gauge zero mode \eqref{prefactor} and the bundles wrapping the base \eqref{basebundles}, we obtain two-dimensional Maxwell theory on $\B$ times a contact term.
After Poisson summation, the partition function of the constant electric field on $\B$ is
\begin{equation}\label{ZE}
Z_E = \sum_{E \in q_\B \mathbb{Z}} e^{-\frac{1}{2} \Vol(\B) E^2},
\end{equation}
while the two-dimensional contact term is given by
\begin{equation} \label{morecontact}
\left( \frac{q_\B}{\sqrt{2 \pi}} \right)^{\chi(\B)}.
\end{equation} 
When we vary the conical angle $\beta$, the volume of $\B$ is proportional to $\beta$, and so the first factor \eqref{ZE} takes the form of a canonical partition function. 
The energy levels are precisely those of the quantized electric field $E$ on the base, for which the geometric entropy \eqref{conical} gives the entanglement entropy \cite{Donnelly2012}.

The bundles polarized along the fiber \eqref{fiberbundles} describe quantized magnetic fields wrapping the fiber directions.  
This contribution is already expressed as a canonical partition function, similar to \eqref{ZE} except that their values are quantized on the lattice $ \frac{2\pi}{q} H^2(\F, \mathbb{Z})$.
Therefore there is no contact term coming from these ``magnetic'' two-dimensional Maxwell fields.

With all the local and topological degrees of freedom accounted for, we are left with the contact term that is the product of \eqref{contact} and \eqref{morecontact}:
\begin{equation} \label{fullcontact}
Z_\chi = \left[ \frac{\sqrt{2 \pi V_\F}}{q} \det{}' (\Delta_0^\F)^{-1/2} \right]^{-\chi(\B)}.
\end{equation}
The geometry of the entangling surface consists of $\chi(\B)$ copies of $\F$, so that $Z_\chi$ is the partition function of a scalar field localized on the entangling surface. However, the sign in the exponent is opposite that of an ordinary bosonic scalar field.
The leading order divergence in the contact term agrees with the expression found in \cite{Kabat1995}, which leads to negative entropy when regulated by heat kernel methods.



\section{Interpreting the contact term} \label{sec:geo}

Given the Kaluza-Klein reduction of Maxwell theory it is now straightforward to calculate the geometric entropy via the conical variation \eqref{conical}.  
Since the partition function is expressed as a product, it is sufficient to calculate the entropy associated to each set of modes separately.
We can then ask whether each individual factor can be given a statistical interpretation.
The local degrees of freedom of Maxwell theory appear after Kaluza-Klein reduction as a tower of free minimally coupled scalar fields.
Because the scalars are minimally coupled, the geometric entropy yields exactly the entanglement entropy, which is well-known for free scalar theories (see e.g. \cite{Casini2009}).

Since the contact term is independent of $\beta$, its contribution to the entanglement entropy is just $S_\chi = \ln Z_\chi$.  The leading order area-law contribution is like the cosmological constant induced by $-1$ scalar fields living on the entangling surface.  In a heat kernel regulator, the negative entropy in fact overwhelms the positive sources of entropy in dimensions $D < 8$.
Thus the ghost scalar fields in the contact term $Z_\chi$ \cite{Kabat1995} render the total entropy negative, and have no obvious statistical interpretation in terms of the actual positive degrees of freedom.  

One might think that this negativity is of little consequence since (in $D > 2$) the area law divergence is a power-law divergence, and the coefficient of power law divergences are nonuniversal.  However, the KK reduction makes it clear that $\ln Z_\chi$ also contributes to the logarithmic divergence (in even dimensions) and to the nonlocal finite piece of the geometric entropy $S$.  This is clear from the fact that it is proportional to the effective action of a $D - 2$ dimensional scalar field.  Thus the problem is \emph{not} an artifact of the renormalization scheme, and cannot be safely removed from the partition function without consequence.\footnote{Among other things, this would make $Z$ no longer invariant under exchanging the roles of $\B$ and a 2D factor manifold of $\F$.}


While subleading terms of divergent quantities can be negative, the negativity of the Kabat term was only a symptom of a deeper concern: is there a statistical mechanical interpretation for these extra terms?  Using the same methods as in \cite{Donnelly2014b}, we will show that the answer is yes.

In section \ref{sec:edge} we will calculate the partition function in a new way, which makes the statistical origin of this contact term clear.  It turns out to be related to the phenomenon of ``edge modes'', new degrees of freedom that appear when restricting a gauge theory to a region with boundary.
The contribution of the electric and magnetic 2D Maxwell fields have a state-counting interpretation related to edge modes, as elucidated in \cite{Donnelly2012,Gromov2014,Donnelly2014a}.  We wish to show that the contact term $Z_\chi$ can \emph{also} be given a statistical interpretation in terms of these edge modes.

To do this, we will regulate the theory using an t'Hooft brick wall at a proper distance $\epsilon$ just outside the entangling surface.  To so so we have to choose boundary conditions at the brick wall, and the key criterion is that they must not affect the physics far from the wall.
However, neither of the standard boundary conditions for Maxwell fields have this property.  If we were to impose electric conducting boundary conditions,\footnote{also known as ``relative'' boundary conditions}
\begin{equation}
\,\,
\alpha = 0, \qquad A_i = 0, 
\qquad \nabla_r A_r - K A_r = 0,
\qquad \,\,\,\,\,\,\,\,\, (r = \epsilon)
\end{equation}
(where $i$ points along the brick wall and $r$ is the proper radial distance coordinate, and $K = K_{ij} g^{ij}$ is the trace of the extrinsic curvature), we would find that there can be no magnetic flux $F_{ij}$ through the entangling surface.  On the other hand, if we imposed magnetic conducting boundary conditions:\footnote{also known as ``absolute'' boundary conditions}
\begin{equation}\label{relative}
\nabla_r \alpha = 0, \qquad A_r = 0, 
\qquad \nabla_r A_i - K_{ij} A^j = 0, 
\qquad (r = \epsilon)
\end{equation}
then the field strength satisfies
\begin{equation}\label{magnetic}
\qquad \qquad
F_{ir} = 0, \qquad \partial_r F_{ij} = 0,
\qquad \qquad \qquad \qquad (r = \epsilon)
\end{equation}
and so we would find that there is no electric flux $E_\perp = F_{r \hat{\tau}}$ (where $\hat{\tau}$ is the unit angular direction around the brick wall in an orthonormal coordinate system).  But in reality the entangling surface is not a physical barrier, so both kinds of flux are allowed.  Thus neither of these boundary conditions are acceptable.

Our solution will be to impose the magnetic boundary conditions with an arbitrary choice of $E_\perp$, by replacing the last equation of \eqref{relative} with
\begin{equation}
\nabla_r A_i - K_{ij} A^j = \hat{\tau}_i E_\perp(\F),
\qquad (r = \epsilon).
\end{equation}
Then we will compensate by doing an explicit path integral over all possible choices of $E_\perp$.  This allows for both electric and magnetic fluxes through the entangling surface.  We will define $Z_\text{bulk}$ as the partition function of the bulk region outside the magnetically conducting brick wall (with $E_\perp = 0$), and $Z_\text{edge}$ as the correction coming from the path integral over the edge modes; thus the total partition function with the wall is
\begin{equation}\label{be}
Z = Z_\text{bulk} Z_\text{edge}.
\end{equation}
Below, we will prove that \eqref{be} agrees with the partition function with no brick wall, even though $\chi(\B) = 0$ for the brick wall system (since $\chi(S_1) = 0$) so that the contact contribution of \eqref{fullcontact} does not contribute.  Nevertheless the contact term still arises in a different way, from the edge mode contribution.


It is almost (but not quite) true that the edge modes $Z_\mathrm{edge}$ give rise to the contact term $Z_\chi$ \eqref{fullcontact} together with the sum over the constant mode of the electric flux $Z_E$ \eqref{ZE}.  There is also an additional term $Z_{D/N}$ coming from the difference between Neumann and Dirichlet boundary conditions for scalars on the brick wall.

In the case of a scalar field it is best to impose Neumann boundary conditions, because this changes the field as little as possible far from the entangling surface.
In the limit $\epsilon \to 0$, there is no effect on the partition function of a scalar field other than through local power law divergences, which are not universal \cite{Donnelly2050}.

On the other hand, for massive scalar fields the Dirichlet boundary conditions have some additional subtleties, including UV divergences of the form $\ln \ln \epsilon^{-1}$ in the entanglement entropy \cite{Donnelly2050}.  This means that the entanglement entropy of the Dirichlet scalar does not quite correspond to the geometric entropy without any brick wall. 

In the case of the Maxwell field coupled to a magnetically conducting brick wall, the tower of scalar fields obtained by KK reduction have Neumann boundary conditions, as can be seen from Eq. \eqref{magnetic} and the fact that $\phi_\text{KK}$ is proportional to the magnetic field $F_{ij}$ (or to $A_i$ in the massless case.)  

But the tower of scalar fields dual to the Proca fields have \emph{Dirichlet} boundary conditions.    This can be seen by substituting the Proca-scalar duality relation $^{2}F_{ab} \epsilon^{ab}/2 = m\phi_\mathrm{dual}$ (\ref{dualize}) into Eq. \eqref{magnetic}, where $^{2}F_{ab}$ is the KK reduced 2-dimensional field strength, which is proportional to the $D$-dimensional field strength $F_{ab}$ polarized along the base $\B$.  

If all of the scalars had Neumann boundary conditions, then because they form a complete set of modes of $\Delta_1^F$, the effects of imposing the brick wall would add up to a contribution which is local along the fiber $\F$.  This means that they could be absorbed into nonuniversal local counterterm on the brick wall.  But in fact, some of the scalar fields have Dirichlet boundary conditions, and we must take this into account.

Let us define $Z_{D/N}$ as the ratio of the partition function for the $\phi_\text{dual}$ modes with Dirichlet boundary conditions, compared to the partition function of these same $\phi_\text{dual}$ modes but with Neumann boundary conditions.  Then the partition function in the presence of a magnetically conducting brick wall is
\begin{equation}
Z_\text{bulk} = Z_B Z_\text{scalars} Z_{D/N}, 
\end{equation}
where the $Z_\mathrm{scalars}$ term refers to the full tower of all scalar fields $\phi_\text{KK}, \phi_\text{dual}$ with Neumann boundary conditions, which is equivalent to no brick wall.

In section \ref{sec:edge} we will show that
\begin{equation}\label{itworks}
Z_\text{edge} Z_{D/N} = Z_\chi Z_E
\end{equation}
which implies that
\begin{equation}
Z = Z_\text{scalars} Z_E Z_B Z_\chi = Z_\text{bulk} Z_\text{edge}
\end{equation}
where the first expression for $Z$ is the geometric partition function with no brick wall, and the second is the brick wall plus edge modes.  Thus we will find an exact agreement between the two partition functions.  Since the corresponding entropies $S_\text{bulk}$ and $S_\text{edge}$ both have a statistical interpretation, the entire entropy has a statistical explanation.

Thus we have explained Kabat's contact term in terms of the statistical mechanics of edge modes, without needing to appeal to the negative entropy ghosts.  The reason why Kabat obtained a negative leading contribution to the entropy will be discussed in section \ref{sec:disc}.

\section{Dirichlet versus Neumann} \label{sec:dn}

In this section we calculate $Z_{D/N}$, which is the ratio of the Dirichlet and Neumann partition functions for the dual scalar modes, which are massive fields on the base $\B$. Recall that there is one dual scalar mode for every nonzero scalar mode on the fiber $\F$.


Let us consider the partition function $Z$ for the manifold $\B_\beta$, where $\B_\beta$ is the conical manifold with angle $\beta$ going around the entangling surface, used to calculate the geometric entropy for some particular field of mass $m$ in \eqref{conical}.  As we zoom in on the entangling surface in the base $\B$, it is approximated by a cone with a small disk of radius $\epsilon$ cut out of the tip, on which we put either Dirichlet or Neumann boundary conditions. 


Consider radial evolution outward from the brick wall in the coordinate $\rho = \ln r$.  We can most easily analyse this problem by doing an exponential conformal transformation with Weyl scaling $\Omega = 1/r = e^{-\rho}$ in order to transform the plane into the $(\tau, \rho)$ coordinate system, with the Cartesian metric $ds^2 = d\rho^2 + d\beta^2$.  The angular coordinate $\tau \in [0,\beta)$ remains periodic, while $\rho \in [\ln \epsilon, \ln R)$, where $R$ is the characteristic length scale of the manifold $\B$ at which the flat approximation is no longer valid; for example, if $\B$ is a sphere $R \sim r_\text{sphere}$.  In 2 dimensions, the propagator of a minimally coupled scalar field is conformally invariant, while the mass term is not.  Therefore the mass term becomes position dependent:
\begin{equation}
I[\phi] = \int d\tau\,d\rho\,\frac{1}{2}\left[(\nabla_a \phi)^2 + e^{2\rho} m^2 \phi^2\right],
\end{equation}
and we can ignore the mass term as $\rho \to -\infty$.

The mass and/or the curved geometry provides a somewhat fuzzy cutoff on one side of the cylinder, but the precise details will turn out not to matter, so long as we take the order of limits so that the conformally transformed ``distance'' to the brick wall $\epsilon$ is larger than any other scale in the problem.
This is valid if we take the brick wall radius $\epsilon$ to be parametrically small compared to the UV cutoff of the theory which cuts off the contributions from large transverse momentum $m$. 
Hence we wish to analyse the theory on a cylinder of length $\ln (R/\epsilon)$ and periodicity $\beta$. 
If we abuse dimensional analysis by assuming $\ln R,\,\ln m \sim 1$, we may write the length as $\ln (\epsilon^{-1})$.

The important thing to notice is that the theory on the cylinder is massless, but not periodic.  Hence there is an IR divergence in the theory, which manifests as the absence of a mass gap on the cylinder when evolving along the $\rho$ direction.

The modes with nonzero $\tau$-momentum \emph{are} gapped, since they correspond to harmonic oscillators. 
Any excitiations of these modes due to the boundary condition decay rapidly away from the entangling surface, so that their contribution is purely local.
This simply shifts the nonuniversal power law contributions to the entropy.

But the zero modes $\phi = \phi_0(\rho)$, which are constant in the $\tau$ direction, are not gapped.  This system corresponds to a free particle whose ``position'' $x = \phi_0/\sqrt{\beta}$ is the canonically normalized zero mode of the field.  Under $\rho$-evolution, the wavefunction of the particle spreads out as a Gaussian.  If we Fourier transform to the momentum $p$ (which is continuous since $\phi_0$ of the scalar field is not periodic), the wavefunction evolves from one end of the cylider to the other like 
\begin{equation}
\Psi(\rho_\text{final}) 
= e^{-\frac{1}{2} \eta p^2} \Psi(\rho_\text{initial})
\end{equation}
where $\eta = \ln(\epsilon^{-1})/\beta$ is the length of the cylinder measured in units of its width.

For Neumann boundary conditions, the wave function is initially a $p=0$ eigenstate $\psi_N(p) = \delta(p)$, and so it is invariant under the radial evolution.  On the other hand, the Dirichlet wave function is initially an $x=0$ eigenstate $\psi_D(p) = (2 \pi)^{-1/2}$, and under radial evolution evolves to
\begin{equation}
(e^{-\frac12 \eta p^2}\psi_D)(p) = (2 \pi)^{-1/2} e^{-\tfrac12 \eta p^2} \underset{\epsilon \to 0}{\longrightarrow} \eta^{-1/2} \delta(p).
\end{equation}
After radial evolution, the Dirichlet wave function approaches the Neumann one times a an extra factor of $\eta^{-1/2}$ (irrespective of the mass $m$ of the field, which makes no difference in the limit $\epsilon \to 0$).

It is interesting to note that because the log which appears in the geometric entropy formula \eqref{conical} stacks onto the log inside $\eta$ coming from the conformal transformation to the cylinder, the entropy of the 2D Dirichlet massive scalar on an interval with two endpoints ends up having a surprising divergence structure:
\begin{equation}
S = \frac{1}{3} \ln \epsilon^{-1} 
- \ln \ln \epsilon^{-1}
+ \text{finite},
\end{equation}
where the first term is the normal log divergence, which in this calculation comes from the Casmir energy of the cylinder vacuum, and the second term is a log of a log.  We plan to address the significance of this term for scalar fields in another paper \cite{Donnelly2050}.  For Maxwell fields with our choice of boundary conditions, this peculiar term will end up cancelling with another term coming from edge modes (section \ref{sec:edge}).

We conclude that imposing Dirichlet boundary conditions leads to an extra factor of $\eta^{-1/2}$ 
for each of the modes of $\F$, except the zero mode.  Since there are $\chi$ points on the entangling surface of $\B$, this factor comes in $\chi$ times and thus we may write
\begin{equation}
Z_{D/N} = 
\left[ \det{}'\left(\frac{\ln(\epsilon^{-1})}{\beta}
\Delta_0^\F \right) \middle/
\det{}'\left(\Delta_0^\F \right)\right]^{-\chi/2}.
\end{equation}
Note that, because zeta function regularization is not invariant under multiplying functional determinants term-by-term, we may \emph{not} cancel the factors of $\Delta_0^\F$ in the determinants in the na\"{i}ve way.  We may however rescale the determinants by shifting each term (including the zero mode of $\Delta_0^\F$ by a constant.  Up to an unimportant local anomaly term (which only affects scheme-dependent quantities), we obtain
\begin{equation}\label{ZDN}
Z_{D/N} = \left(\frac{\ln(\epsilon^{-1})}{\beta}\right)^{\chi/2}
\end{equation}
This correction will be important for establishing the equivalence \eqref{itworks} between the entropy calculations with and without the brick wall.

\section{Entanglement entropy and edge modes} \label{sec:edge}

To interpret the contact term as an entanglement entropy, we return to the question of how to define entanglement entropy in a gauge theory.
We first recall the definition of entanglement entropy for Hamiltonian lattice gauge theories used in Ref.~\cite{Donnelly2011}, and then take its continuum limit.

In the Hamiltonian formulation of $U(1)$ lattice gauge theory, a convenient basis for the Hilbert space is the electric field basis.
Each vector in this basis is labelled by a quantized electric flux $E_{vw} \in q \mathbb{Z}$ assigned to each oriented link $(v,w)$ of the lattice.
These must obey Gauss' law, that the electric flux at each vertex is zero, $\sum_{v} E_{vw} = 0$.
The Hilbert space is spanned by superpositions of these electric states.

To define the entanglement entropy on this lattice, we partition the vertices into two sets $A$ and $B$.
The entangling surface then intersects the lattice in some set of edges.
We define the Hilbert space $\mathcal{H}_A$ of region $A$ to be spanned by the electric field states that include all edges that intersect the region, including the boundary.
The physical states, those satisfying Gauss' law, can be identified as states in $\mathcal{H}_A \otimes \mathcal{H}_B$, but they do not span the full tensor product.
The Hilbert space $\mathcal{H}_A \otimes \mathcal{H_B}$ includes states for which the electric flux on the boundary of $A$ does not match with the electric flux on the boundary of $B$.
Although the full Hilbert space $\mathcal{H}$ does not admit a local factorization, we can embed the physics states into the tensor product $\mathcal{H}_A \otimes \mathcal{H}_B$ and then define entanglement entropy normally.

For gauge-invariant states, the reduced density matrix of a region commutes with the operator measuring electric flux through any link on the boundary.
As a result, the density matrix can be split into a direct sum of superselection sectors, each labelled by the configuration of $E_\perp$ on the entangling surface: 
\begin{equation}
\rho = \sum_{E_\perp} p(E_\perp) \rho(E_\perp).
\end{equation}
The resulting entropy is given by
\begin{equation} \label{Sent}
S = -\sum_{E_\perp} p(E_\perp) \ln p (E_\perp) + \sum_{E_\perp} p(E_\perp) S(\rho(E_\perp)).
\end{equation}
We will now define the entanglement entropy as the continuum limit of this expression.

For each distribution of surface charges $E_\perp$, we define the edge mode to be the unique static classical solution of the form $E = \nabla \varphi$ with the boundary condition $\nabla_\perp \varphi = E_\perp$.
Any field configuration can be expressed as a sum of an edge mode and a fluctuation satisfying the magnetic conducting boundary condition $E_\perp = 0$.
Since Maxwell theory is linear, the action for such a configuration is the sum of the on-shell action of the edge mode and the action of the fluctuation.
The term $S(\rho(E_\perp))$ in \eqref{Sent} is therefore independent of $E_\perp$, so it is simply the entropy of the theory with magnetic conducting boundary conditions given by \eqref{magnetic}.

The first term in \eqref{Sent} is the entanglement entropy of the edge modes, and the second is the bulk entropy, so we have
\begin{equation}
S = S_\text{edge} + S_\text{bulk}.
\end{equation}

To calculate $S_\text{edge}$ we define an edge mode partition function by integrating over the edge modes weighted by their on-shell action:
\begin{equation} \label{Zedge}
Z_\text{edge} = \int \D E_\perp e^{-I(E_\perp)}.
\end{equation}
The measure in this expression is taken to be the the continuum limit of the sum in \eqref{Sent}.
We can obtain the entropy of the edge modes $S_\text{edge}$ from the geometric entropy formula \eqref{conical}.


In order to make sense of the formal expression \eqref{Zedge} we will need to introduce a short-distance cutoff near the entangling surface. 
We will do this by introducing a boundary at $r = \epsilon$, similar to the brick wall model of Ref.~\cite{tHooft1985}, except that rather than fixing boundary conditions we sum over all values of the perpendicular electric field $E_\perp$.

In order to compare the entanglement entropy defined in \eqref{Sent} with the geometric entropy, it is sufficient to compare the partition functions at arbitrary $\beta$.
Previously we saw that the geometric partition function (with no brick wall) can be re-expressed in the form
\begin{equation}
Z = Z_\text{bulk} Z_\text{edge}.
\end{equation}
so long as the following identity is true:
\begin{equation}\label{goal}
Z_\text{edge} Z_{D/N} = Z_\chi Z_E
\end{equation}
We now show that these two partition functions are in fact equal, so that the contact term does indeed have a statistical interpretation, as it contributes to the entropy of the edge modes.

In \ref{edge:action} we calculate the Euclidean action for each edge mode appearing in Eq.~\eqref{Zedge}, and show that the partition function is that of a negative scalar field on the entangling surface.
In \ref{edge:measure} we compute the measure appearing in Eq.~\eqref{Zedge} by taking the continuum limit of a lattice regulator.
This gives the appropriate factors of $q$ and $\Vol(\F)$ that appear in the contact term \eqref{fullcontact}, as well as the term $Z_{\D/N}$ in \eqref{goal}.

\subsection{Short-distance expansion} \label{edge:action}

In our product geometry $\M = \B \times \F$, the entangling surface consists of some number of points in $\B$ each with a conical singularity of angle $\beta$.
Let us begin by choosing one of these points around which we will fix polar Riemann normal coordinates with the point at the origin $r=0$.
In these coordinates the metric takes the form
\begin{equation} \label{shortdistance}
ds^2 = dr^2 + N(r)^2 d \tau^2,
\end{equation}
where $\tau$ is $\beta$-periodic and the lapse function is $N(r) = r + \mathcal{O}(r^2)$.
We place a brick wall at $r = \epsilon$, and solve for the classical solution with fixed $E_\perp$ on the brick wall.

The solutions have qualitatively different behavior for the mode of $E_\perp$ constant along $\F$ and for the higher modes.
For the constant mode the solutions have a constant electric field throughout $\B$ whose value is quantized in multiples of the fundamental charge $q_\B$.
This is precisely the contribution to the partition function $Z_E$ \eqref{ZE} coming from the constant electric fields.  We will separate out this contribution and focus on the nonzero modes in what follows.

For the nonzero modes the classical solutions decay rapidly away from the entangling surface.
This is because configurations with electric field lines extending far from the surface  have a large boost energy.
Because these solutions closely hug the entangling surface they have a small action associated with them - and they make a large contribution to the partition function that is local to the entangling surface.
Since the solutions contributing to the sum over edge modes do not extend far from the entangling surface, it will be sufficient to treat each connected component of the entangling surface independently.

For each distribution of surface charge $E_\perp(x)$, we have to find the action of the corresponding classical solution.
Let us expand the vector potential in modes $\psi_n(x)$ of the fiber $\F$ as
\begin{equation}
A = \sum_{n} A_n \varphi_n(r) \psi_n(x) d \tau,
\end{equation} 
We will take $\psi_n$ to be eigenfunctions of the scalar Laplacian on the fiber $\Delta_0^\F \psi_n = \lambda_n \psi_n$, normalized so that $\int_\F \sqrt{g_\F} \, \psi_n^2 = 1$.

In order for $A$ to describe a classical solution, we must have $d \star F = 0$:
\begin{eqnarray}
F &=& (\partial_r \varphi_n) \psi_n \, dr \wedge d \tau + \varphi_n d\psi_n \wedge d\tau \\
\star F &=& N^{-1} \left[ (\partial_r \varphi_n) \psi_n \mathrm{Vol}_\F + \varphi_n \star_\F d \psi_n \wedge dr \right] \\
d \star F &=& (\partial_r N^{-1} \partial_r \varphi_n) \; \psi_n \; dr \wedge \mathrm{Vol}_\F + N^{-1} \varphi_n \; (d \star_\F d \psi_n) \wedge dr \\
&=& (\partial_r N^{-1} \partial_r \varphi_n - N^{-1} \lambda_n \varphi_n) \psi_n dr \wedge \mathrm{Vol}_\F.
\end{eqnarray}
Thus the classical equation of motion reduces to
\begin{equation} \label{EOM}
N \partial_r N^{-1} \partial_r \varphi_n = \lambda_n \varphi_n.
\end{equation}

The on-shell action of this solution can be obtained by using the equation of motion to integrate by parts.
This is analogous to the way in which the electrostatic energy of a system of charges can be expressed in terms of the potential evaluated at the charges.
The on-shell action is given by
\begin{equation} \label{action-onshell}
S 
= \frac12 \int F \wedge \star F 
= \frac12 \int dA \wedge \star F 
= \frac12 \oint A \wedge \star F.
\end{equation}
This depends only on the field values at the brick wall, so we need only find the asymptotic expansion of the solution to \eqref{EOM} for $r \to 0$.

For small $r$, the solutions of \eqref{EOM} have the leading-order asymptotic behavior 
\begin{equation} \label{asymp}
\varphi_n(r) = a + b r^2 \ln r + O(r^2).
\end{equation}
For such a solution, the equation of motion \eqref{EOM} relates the coefficients of the asymptotic expansion as
\begin{equation} \label{ab}
2 b = \lambda_n a.
\end{equation}
This allows us to relate the potential at the brick wall to the normal electric field. 
Let us expand the perpendicular electric field on the boundary as $E_\perp(x) = \sum_n E_n \psi_n(x)$.
The electric field at the brick wall is 
\begin{equation}
E_n = -N^{-1} \partial_r \varphi_n = 2 b \ln (\epsilon^{-1}).
\end{equation}
and the value of the potential at the surface is determined up to terms of order $\epsilon^2$ by $E_n$ as
\begin{equation}
\varphi_n|_{r=0} = a = \frac{2b}{\lambda_n} = \frac{E_n}{\lambda_n \ln (\epsilon^{-1})}.
\end{equation}
Inserting the mode expansion into the on-shell action \eqref{action-onshell} we find that
\begin{equation}
I(E_\perp)
=   \frac12 \sum_n \oint_{\partial \B \times \F} \varphi_n N^{-1} (\partial_r \varphi_n) \; d\tau \wedge \psi_n^2 \mathrm{Vol}_\F 
= \frac12 \sum_n \oint_{\partial \B} \varphi_n N^{-1} (\partial_r \varphi_n) d\tau = 
\frac12 \sum_n \int \varphi_n E_n d\tau.
\end{equation}
From the asymptotic expansion \eqref{asymp}, we find that to leading order in $\epsilon$,
\begin{equation} \label{action}
I(E_\perp) = \sum_n \frac{\beta E_n^2}{2 \lambda_n \ln (\epsilon^{-1})}.
\end{equation}
The eigenvalue $\lambda_n$ appears in the denominator; thus the integral over $E_\perp$ will lead to a functional determinant $\det(\Delta)^{+1/2}$, like the partition function of a negative scalar.

Note that the argument $\epsilon^{-1}$ of the logarithm in \eqref{action} is dimensionful, and needs to be compensated by another dimensionful factor.
This dimensionful factor is determined either by $\lambda_n$, which is related to the transverse wavenumber of the perturbation, or by the length scale $R$ associated to the background.
This is the same behavior as in the calculation of $Z_{D/N}$ in section \ref{sec:dn}.
In the case where $R$ is large we can appeal to the Rindler limit in which $N(r) = r$.
Then the solution to Eq.~\eqref{EOM} is 
\begin{equation}
\varphi_n(r) = r K_1(\sqrt{\lambda_n} r) \sim 1 + \frac{\lambda_n r^2}{2}  \ln (\sqrt{\lambda_n} r) + \mathcal{O}(r^2).
\end{equation}
In this case the dimensionful factor of $\epsilon$ in the action is compensated by $\lambda_n$.
In de Sitter space of radius $R$, one can show that
\begin{equation}
\varphi_n \sim 1 + \frac{\lambda_n r^2}{2} \ln (r / R) + \mathcal{O}(r^2)
\end{equation}
where the exact solution can be written in terms of hypergeometric functions.
Which dimensionful constant compensates for the dimensions of $\epsilon$ depends on whether the given mode is larger or smaller than the length scale of $\B$.
These logarithms appearing in the action lead to $\ln \ln R$ terms in the entanglement entropy, similar to those that appeared for massive scalars in Ref.~\cite{Casini2009}.
These types of terms will be analyzed in more detail in a future work \cite{Donnelly2050}.

\subsection{Functional measure} \label{edge:measure}

We have established that the edge modes take the form of a negative scalar determinant on the entangling surface, which agrees with the contact term up to a constant prefactor.
In order to match this prefactor, we have to carefully define the path integral measure $\D E_\perp$. 
We do this by taking the continuum limit of the discrete measure on the lattice.
Consider a discrete set of $N$ boundary points $\{ x_i \}$ in $\F$, representing the points at which the links of the lattice pierce the entangling surface.
We imagine the surface is tesselated so that each of these points is assigned a volume $\Vol(\F)/N$.
In the lattice theory, the integrated electric flux through each point
is quantized in units of $q$.
The sum over these discrete values defines the lattice measure
\begin{equation} \label{discrete-measure}
\int \D E_\perp \equiv \prod_{i} \sum_{E_\perp(x_i) \in \frac{q N}{\Vol(\F)} \mathbb{Z}} \delta \left (\sum_i E_\perp(x_i), 0 \right).
\end{equation}
The Kronecker $\delta$ ensures that we sum only over configurations such that the total flux through the entangling surface vanishes; this is because we have already accounted for solutions for which the flux through the entangling surface is constant as part of $Z_E$.

In order to carry out the integration, we can change variables from $E_\perp(x_i)$ to the coefficients $E_n$ of a mode expansion via
\begin{equation}
E_\perp(x) = \sum_n E_n \psi_n(x).
\end{equation}
To take the continuum limit we make the replacement 
\begin{equation}
\sum_{E_\perp(x) \in \frac{qN}{\Vol(\F)} \mathbb{Z}} \to \frac{\Vol(\F)}{qN}\int \; d E_\perp(x).
\end{equation}
Expressed in terms of the mode expansion, the measure \eqref{discrete-measure} (without the delta function) is
\begin{equation}
\mathcal{D} E_\perp = \prod_{n} \frac{1}{q} \sqrt \frac{\Vol(\F)}{N} d E_n.
\end{equation}
This can be seen by comparing the calculation of $\int \mathcal{D}E_\perp \exp \left( -\int_\F \sqrt{g_\F} E_\perp^2(x) \right)$ in the mode expansion and in the continuum limit of \eqref{discrete-measure}.

We now have to restrict to those field configurations such that the total electric flux through the surface vanishes.
This can be obtained by inserting the $\delta$-function:
\begin{equation} \label{delta}
\delta \left( \frac{\Vol(\F)}{q N} \sum_i E_\perp(x_i) \right) = \delta\left( \frac{\sqrt{\Vol(\F)}}{q} E_0 \right)
\end{equation}
where $E_0$ is the coefficient of the constant zero mode.
Performing the zero mode integral with the help of Eq.~\eqref{delta} we are left with the path integral measure
\begin{equation} \label{measure}
\D E_\perp = \frac{1}{\sqrt{N}} \prod_{n > 0} \frac{1}{q} \sqrt{\frac{\Vol(\F)}{N}} d E_n.
\end{equation}

Now we can carry out the path integral using the measure \eqref{measure} and the action \eqref{action}, with the result
\begin{equation} \label{Zedge-unscaled}
Z_\text{edge} = \frac{Z_E}{\sqrt{N}} \prod_{n > 0} \frac{1}{q} \sqrt{\frac{\Vol(\F)}{N}} \left( \frac{2 \pi \ln (\epsilon^{-1}) \lambda}{\beta} \right)^{1/2}
= \frac{Z_E}{\sqrt{N}} \det{}' \left( \frac{2 \pi \Vol(\F) \ln(\epsilon^{-1})}{q^2 N \beta} \Delta_0^\F \right)^{1/2}
\end{equation}
including the factor $Z_E$ coming from the quantized constant edge mode.
We can now rescale this formula using the zeta function identity \eqref{zeta}: $\det{}' (a \Delta) = a^{\zeta(\Delta,0)} \det{}' (\Delta)$.
Up to an anomaly term, $\zeta(\Delta_0^\F,0) = - \dim \ker (\Delta_0^\F) = -1$.
After rescaling, and taking into account that the entangling surface consists of $\chi(B)$ components, the edge mode partition function is
\begin{equation}\label{finaledge}
Z_\text{edge} = Z_E \left( \sqrt{\frac{\ln (\epsilon^{-1})}{\beta}} \frac{\sqrt{2 \pi \Vol(\F)}}{q} \det(\Delta_0^\F)^{-1/2} \right)^{-\chi(\B)}.
\end{equation}
We see that the partition function of the edge states agrees with the result from the contact term \eqref{fullcontact}, up to the factors $Z_E$ \eqref{ZE} and $Z_{D/N} = (\ln(\epsilon^{-1})/ \beta)^{\chi/2}$ \eqref{ZDN}, which appear exactly as needed \eqref{goal} to make the geometric and entanglement entropies agree.  Thus we have provided the geometric entropy with a manifestly statistical interpretation.
    
\section{Logarithmic divergence in four dimensions}  \label{sec:log}

The case of four dimensions is special, because four-dimensional Maxwell theory is conformal.
In a conformal field theory, the entanglement entropy of a spherical region has a universal logarithmic divergence, whose coefficient is related to the trace anomaly \cite{Solodukhin2008,Casini2011}.

Four dimensions is also interesting because of the connection to AdS/CFT.
In a conformal theory with a holographic dual described by Einstein gravity, the Ryu-Takayanagi (RT) formula relates the entanglement entropy to the area of a minimal surface that extends into the bulk \cite{Ryu2006a}.
The logarithmic term in this entropy is related to the holographic trace anomaly \cite{Henningson1998}, which is protected by supersymmetry \cite{Petkou1999}.
Thus, although the RT formula becomes tractable only at strong coupling, there is a universal part that should agree at strong and weak coupling.
Thus the entanglement entropy gives us a possible check on the RT formula that can be carried out at weak coupling.

While the RT formula has passed many nontrivial checks \cite{Nishioka:2009un}, given the subtlety of entanglement entropy in gauge theories, there is still the question of \emph{which} entropy RT calculates, given that there are multiple possible candidates for the role of ``entanglement entropy'' \cite{Casini2013}.  The derivations of the RT formula \cite{Fursaev:2006ih, Casini2011, Lewkowycz:2013nqa} involve the geometric entropy (or the related replica trick \cite{Calabrese2009}) and therefore one expects that RT calculates the geometric entropy.  We shall show below that at weak coupling, for Maxwell fields, this geometric entropy includes a contribution from edge modes, as needed to agree with the trace anomaly.  Since RT also agrees with the trace anomaly, we expect that the RT entanglement entropy of strongly coupled Yang-Mills theory already includes an edge mode contribution, but we will only perform the edge mode calculation at weak coupling.

Consider the entanglement entropy of a ball of radius $r$ in $3+1$ dimensions.
Using conformal symmetry, the entanglement entropy can be equivalently expressed as the thermal entropy in hyperbolic space $\mathbb{R} \times H^3$, or as de Sitter entropy of the static patch of de Sitter space.
The logarithmic divergence in the entanglement entropy is related to the $a$-type trace anomaly \cite{Casini2011},
\begin{equation}
\left \langle T^a_a \right \rangle = -\frac{a}{16 \pi^2} E_4 + \frac{c}{16 \pi^2} C_{abcd} C^{abcd},
\end{equation}
where $E_4 = R_{abcd}R^{abcd} - 4 R_{ab} R^{ab} + R^2$ is the $4$-dimensional Euler density, whose integral is $\int \! \sqrt{g} \, E_4 = 32 \pi^2 \chi$, and $C_{abcd}$ is the Weyl tensor.
The logarithmic divergence in the entropy is given by:
\begin{equation}
S_\text{anom} \sim 4 a \ln (\epsilon/r),
\end{equation}
where $\sim$ denotes agreement of logarithmic divergences, and $\epsilon$ is an ultraviolet cutoff length.
For a theory of $n_0$ scalars, $n_{1/2}$ Dirac fermions and $n_1$ gauge fields the trace anomaly predicts a geometric entropy \cite{Birrell1982}
\begin{equation} \label{Sanomaly}
S_{\text{anom}} \sim \frac{1}{90} \left( n_0 + \frac{11}{2} n_{1/2} + 62 n_1 \right) \ln (\epsilon/r).
\end{equation}

Ref.~\cite{Dowker2010} calculated the entanglement entropy of a free theory of $n_0$ scalars, $n_{1/2}$ Dirac fermions and $n_1$ gauge fields.
The logarithmic divergence in the entropy was found to be:
\begin{equation} \label{Sthermal}
S_{\text{thermal}} \sim \frac{1}{90} \left( n_0 + \frac{11}{2} n_{1/2} + 32 n_1 \right) \ln (\epsilon /r).
\end{equation}
This result was obtained by making a conformal transformation to de Sitter space, calculating the thermal entropy as a function of temperature, and then integrating the first law of thermodynamics.
Comparing \eqref{Sanomaly} with \eqref{Sthermal}, we see that the thermal entropy of the scalar and spinor fields agree with the corresponding trace anomalies, but for the gauge fields the two differ.  Since the coefficients of log divergences are expected to be universal (i.e. independent of the regulator scheme), this discrepency cannot be attributed simply to the choice of regulator.

However, the discrepency can be resolved by including the entanglement of edge modes.  When integrating the first law,  \cite{Dowker2010} assumed that the entropy vanishes in the zero temperature limit.
That would have been the case if we had kept the finite lattice spacing (introduced temporarily in section \ref{sec:edge}), but it is not true for the edge modes in the continuum, even at finite brick wall radius, because there are a continuum of edge modes at any nonzero temperature.  There is a contribution which is independent of $\beta$, and therefore does not vanish in the $\beta \to +\infty$ limit.
This divergent contribution to the entropy is missed by the thermodynamic calculation of Ref.~\cite{Dowker2010}.

The result \eqref{Sthermal} can also be found following \cite{Eling2013,Huang2014b} by calculating the thermal entropy density on $H^3$ and multiplying by the regularized volume.
However such a procedure misses any non-extensive contributions coming from boundary effects, such as the edge mode contribution of section \ref{sec:edge}.

We now calculate the entropy of the edge modes on hyperbolic space and show that when added to the thermal result \eqref{Sthermal}, the result is in agreement with the trace anomaly \eqref{Sanomaly}.
The argument proceeds almost identically to section \ref{sec:edge}, except that the manifold is no longer a product manifold.
Instead we use conformal symmetry to map the entanglement entropy of a sphere to the thermal entropy on the static universe $S^1 \times H^3$, for which the metric is
\begin{equation}
ds^2 = d \tau^2 + du^2 + \sinh(u)^2 d\Omega_2^2.
\end{equation}
Under this transformation the entangling surface is mapped to $u = \infty$, so that the brick wall at $r = \epsilon$ is mapped to
\begin{equation}
u_\text{max} = -\ln \frac{\epsilon}{2r}.
\end{equation} 

To find the edge modes we fix the electric flux $E_\perp$ at $u = u_\text{max}$, and solve for the potential $\varphi$ in the interior:
\begin{equation} \label{Laplace}
\nabla^2 \varphi = 0, \qquad \partial_u \varphi\,|_{u = u_\text{max}} = E_\perp.
\end{equation}
We can expand $E_\perp$ and $\varphi$ in spherical harmonics: 
\begin{eqnarray}
E_\perp(\theta,\phi) &=&  \sum_{\ell > 0,m} E_{\ell m} Y_{\ell m}(\theta, \phi), \\
\varphi(u,\theta,\phi) &=& \sum_{\ell > 0,m} \varphi_{\ell,m}(u) Y_{\ell m}(\theta, \phi).
\end{eqnarray}
Equation \eqref{Laplace} then becomes
\begin{equation}
\partial_u^2 \varphi_{\ell,m}(u) + 2 \tanh(u) \partial_u \varphi_{\ell,m}(u) - \sinh(u)^{-2} \ell (\ell+1) \varphi_{\ell,m}(u) = 0.
\end{equation}
whose solution is
\begin{equation}
\varphi_{\ell,m}(u) \propto \tanh (u)^\ell \; {}_2 F_1 \left( \tfrac{\ell}{2}, \tfrac{\ell+1}{2}, \tfrac{2 \ell + 3}{2}, \tanh(u)^2 \right)
\end{equation}
where ${}_2 F_1(a,b;c;z)$ is the hypergeometric function. 


To find the action, we can use the asymptotic expansion of $\varphi$, given by
\begin{equation}
\varphi_{\ell,m} \propto 1 - 2 \ell (\ell + 1) u e^{-2 u} + \mathcal{O}(e^{-u})
\end{equation}
This gives the value of $\varphi$ at $u = u_\text{max}$ to leading order in $\epsilon$,
\begin{equation}
\varphi_{\ell m} = \frac{\epsilon^2 \ln \epsilon^{-1}}{\ell (\ell + 1) r^2} E_{\ell m}.
\end{equation}
The Euclidean action is then given by
\begin{equation} \label{action-h3}
I(E_\perp) = \oint E_\perp(\theta, \phi) \varphi(\theta, \phi) = \sum_{\ell,m} \beta \frac{\epsilon^2 \ln \epsilon^{-1}}{\ell (\ell + 1) r^2} E_{\ell m}^2.
\end{equation}
Since we are interested in the logarithmic divergence, we can ignore the $\ell$-independent factors in the action and the measure.

We recognize the factor $\ell (\ell + 1)$ in \eqref{action-h3} as the spectrum of the Laplacian on the sphere.
Since it appears in the denominator, it leads to an edge mode partition function equal to that of a wrong-sign scalar field, exactly as in \ref{sec:edge}.
The logarithmic divergence of this partition function is given by the well-known trace anomaly of a two-dimensional scalar field,
\begin{equation}
\left \langle T^a_a \right \rangle = - \frac{R}{24 \pi}.
\end{equation}
This determines the logarithmic divergence in the partition function of the edge modes, and hence the entropy
\begin{equation}
S_\text{edge} \sim \ln Z_\text{edge} \sim \frac13 \ln (\epsilon).
\end{equation}
Thus the thermal entropy \eqref{Sthermal} should be supplemented by an edge mode contribution for each of the $n_1$ gauge fields.
This result agrees with the result of the trace anomaly \eqref{Sanomaly}.

\section{Discussion} \label{sec:disc}

We have shown that in Maxwell theory, the geometric entropy calculated by Euclidean path integral methods agrees with the entanglement entropy using a brick wall regulator plus edge modes.  Since the latter system manifestly has a statistical interpretation, the former does as well.  The contact interaction, which in the Euclidean path integral comes from nonminimal coupling to curvature, is equivalent to the entanglement entropy of the edge modes (up to the correction described in section \ref{sec:dn}).
This resolves a longstanding puzzle about the interpretation of the contact term.

To leading order, the contact term takes the form of a negative scalar field confined to the entangling surface.
The contact term affects the universal subleading divergences in the entanglement entropy, which are the coefficient of the logarithmic divergence in even dimensions and the constant in odd dimensions.
Since the entangling surface has codimension two, the edge modes have subleading divergences of the same order, and hence contribute to these universal terms.
In section \ref{sec:log} we examined these divergences in the case of 3+1 dimensions, resolving an apparent discrepancy first noted in Ref.~\cite{Dowker2010} between the entanglement entropy and the trace anomaly.

In sections \ref{sec:kk} through \ref{sec:dn} we have assumed a special class of geometries of the form $\B \times \F$.
The main role of this assumption was to be able to isolate the contact term contribution to the partition function by expressing all the local degrees of freedom as scalar fields.
In more general geometries, it is not clear whether such a division is possible.
However, the division of the entropy into an edge mode contribution and a bulk contribution \eqref{Sent} is more general and does not depend on the background geometry.
It is straightforward to generalize the edge mode calculation to geometries with a rotational symmetry around the entangling surface, as was done in section \ref{sec:log} in the special case of hyperbolic space.
This symmetry allowed us to use the geometric entropy formula \eqref{conical}.
To calculate entanglement entropy for more general regions by Euclidean methods one should instead use the replica trick \cite{Calabrese2009}.
In this more general setting, the entangling surface may have nonzero extrinsic curvature that must be taken into account in the short distance expansion near the entangling surface \eqref{shortdistance}.
This would allow us to determine the dependence of the entanglement entropy on the extrinsic geometry of the entangling surface, to compare to \cite{Solodukhin2008,Fursaev:2013fta}.

We used a lattice regulator to define the measure of the path integral over edge modes, and we expect that we would obtain similar results if we had calculated the entropy entirely on the lattice (perhaps using the replica trick).  The leading order area law divergence of the von Neumann entropy on the lattice should be manifestly positive, but we expect that the log divergences should agree with the continuum result.

In this article we have only considered the case of abelian gauge fields; it would be interesting to extend the results to the nonabelian case.
For some purposes $\text{SU}(N)$ gauge fields can be treated in the weakly coupled limit as if it were $N^2-1$ decoupled copies of the U(1) abelian theory, modulo zero mode issues related to the different shape and topology of the gauge group.
Thus one would expect at weak coupling the entropy should grow like $N^2$.  If, following Ref.~\cite{Casini2013}, we allow only entanglement between gauge invariant operators this seems unlikely to be the case, since the number of gauge-invariant operators grows more slowly than the number of local degrees of freedom.
In the definition of entanglement entropy proposed for lattice gauge theories in Ref.~\cite{Donnelly2011} there is an additional term in the entropy that grows with the dimension of the gauge group, and which might correct for this deficiency.
It remains to be seen whether either of these definitions of the entropy has the anticipated quadratic scaling with $N$.

A comment is warranted about the negative sign in the exponent of the contact term in the heat kernel; this is the phenomenon that led to the conclusion in Ref.~\cite{Kabat1995} that the Maxwell field has negative entropy.
It seems surprising that one could arrive at a negative entropy starting from a manifestly positive lattice expression. 
The negative result arises in taking the continuum limit of this discrete expression.
To see how such a term arises, consider taking the continuum limit of the entropy $S = - \sum p \ln p$ over a discrete set of values with spacing $a$.
Now consider the limit $a \to 0$, with the probability density $\rho = p/a$ held fixed, so that $S = -\sum p \ln p \to -\int \rho \ln \rho - \ln a$.
The former expression is the continuous entropy, which becomes negative for tightly spaced distributions, but this negative term is always compensated by the positive divergent piece $- \ln a$. 
However when we sum over modes the continuous entropy becomes increasingly negative for higher modes as the action increases with the value of the electric field. 
The $-\ln a$ term, on the other hand, is the same for each mode and can be absorbed into a rescaling of the fields or as a local counterterm.
Thus one arrives in the continuum at an expression that appears formally negative.

Zeta function regularization eliminates all power law divergences, but one can retain the power law divergences by instead inserting a UV cutoff $\epsilon$ directly into the heat kernel expression for the partition function of a scalar field, as done by Kabat \cite{Kabat1995}:
\begin{equation} \label{heatkernel}
\ln Z = \frac12 \int_{\epsilon^2}^\infty \frac{\tr e^{-s \Delta}}{s} ds = \cdots -\tfrac12 \ln \det \Delta,
\end{equation}
where the ``$\cdots$'' represent power-law divergences \cite{Hawking1976} (including a constant in even dimensions).  Note that even a single mode produces a $\ln \epsilon^{-1}$ divergence when inserted into this expression.  This occurs because the path integral measure for each scalar mode requires a dimensionful factor $\mu$ to be inserted for it to be well defined.  In the heat kernel regulator, the convention is that we choose $\mu = \epsilon^{-1}$, but as a consequence even a single mode appears to be UV divergent.  Because of this convention, one finds that the leading order cosmological constant divergence of the scalar field is positive.  If we had instead imposed a hard momentum cutoff $\Lambda$, while leaving $\mu$ fixed in the $\Lambda \to \infty$ limit, the leading order cosmological constant becomes negative.  Since the edge mode contribution is like minus one scalar field, a hard cutoff on the edge modes would produce a positive leading contribution to the entropy.  This illustrates the dangers in assigning too much significance to the divergent terms generated by the heat kernel method.

Thus we see that the negativity of the Kabat contact term is an artifact of the particular regulator scheme he used.  However, the ambiguity of sign is only possible in the first place because the edge modes are continuous and need a path integral measure $\D E_\perp$ to be well defined.


A similar negative contact term appears in the heat kernel for gravitons \cite{Solodukhin2011, Fursaev1996}.  Presumably the graviton contact term can also be understood in terms of edge modes due to gravitational constraints.  However, there are many additional conceptual problems involved in defining the entanglement entropy of graviton fields (reviewed in Appendix A of \cite{Bousso:2015mna}).

The entanglement entropy of quantum fields on a black hole background provides an additive contribution to the total entropy (references in \cite{Bousso:2015mna}), and may conceivably explain it all, when integrated up to the Planck scale \cite{Susskind1994, Jacobson1994}.  In the limit of a large black hole, we may write the entropy as
\begin{equation}
S_{BH} = \frac{A}{4G} 
= \frac{A}{4G(\Lambda)} + S_\mathrm{matter},
\end{equation}
where $G(\Lambda)$ is the effective value of the Newton's constant at the energy scale $\Lambda$.  If $S_\mathrm{matter}$ has a positive divergence, this leads to screening of $G$.

On the other hand, if gravitons and gauge fields provide a negative divergence in $S_\mathrm{matter}$, this leads to antiscreening of Newton's constant $G$.  This suggests that $G$ may flow to a UV fixed point at positive values of $G$.  This is the asymptotic safety scenario proposed in \cite{Weinberg1979, Smolin:1979uz} (see \cite{Niedermaier:2006wt, Reuter:2012id} for reviews).  However, the RG flow of $G$, being a power law divergence in $D > 2$, is a scheme-dependent quantity, raising questions about whether this provides meaningful evidence about the consistency of the theory in the deep UV \cite{Anber:2011ut}.  In a different scheme where the entropy is inherently positive, such as the lattice regulator, any UV fixed point would need to be at negative values of $G$.

Moreover, a negative $S_\mathrm{matter}$ would be in some tension with the expected positivity of entropy from the statistical mechanics of black holes.  It is expected that the black hole entropy $A/4G$ has a microscopic state counting interpretation in terms of a true von Neumann entropy $-\text{tr}(\rho \ln \rho)$ of a discrete number of states.  
For this to be true, the theory of quantum gravity must somehow discretize the continuous $E_\perp$ degrees of freedom, much as the lattice regulator does.  
If this is the right way to think about black hole entropy, it would seem to follow that $S_\mathrm{edge}$ should be positive in the fundamental theory of quantum gravity.

The contact terms also appear to play a role in making the perturbative one-loop string contribution to the entanglement finite, cutting it off at the string length scale \cite{Susskind1994, Kabat1995, He2014}.  Here the negative and positive divergences exactly cancel.  But since a negative divergent contribution to the black hole entropy makes little sense for the reasons described above, presumably this description is valid only at distances larger than the Planck scale.  We expect that in a nonperturbative definition of string theory, the black hole entropy will be manifestly positive.

\paragraph*{Acknowledgements}

We are grateful for conversations with Ted Jacobson, Don Marolf, Mark Srednicki, Dan Kabat, Sergey Solodukhin, Joe Polchinski, Chris Eling, Ben Michel, Ed Witten, Josh Cooperman, Markus Luty, Xi Dong, Juan Maldacena, Debajyoti Sarkar, Horacio Casini and Ariel Zhitnitsky.
We also acknowledge the hospitality of Perimeter Institute and the KITP while parts of this work were being completed.
W.D. is supported by funds from the University of California.
A.W. is supported by the Institute for Advanced Study, the Simons Foundation, and NSF Grants No. PHY-1205500, PHY11-25915, PHY-1314311.

\bibliographystyle{utphys}
\bibliography{kk}

\end{document}